\documentclass[preprint,showpacs,preprintnumbers,amsmath,amssymb,superscriptaddress]{revtex4}
\include{ams}
\usepackage{amsmath,amssymb,amsthm}
\usepackage{dcolumn}
\usepackage{bm}
\usepackage{graphicx}
\usepackage{float}
\usepackage{subfig}

\begin{document}

 \title{Market inefficiency identified by both single and multiple currency trends}

 \author{T.Tok\'ar} \email{tomastokar@gmail.coms}
 \author{D. Horv\'ath} \email{horvath.denis@gmail.com}
 \affiliation{Sors Research a.s., Stroj\'arensk\'a 3, 040 01 Ko\v{s}ice, Slovak Republic} 

 
 \begin{abstract}

 Many studies have shown that there are good reasons to claim very low predictability of currency returns; 
 nevertheless, the deviations from true randomness exist which have potential predictive and 
 prognostic power~[J.James, Quantitative finance 3 (2003) C75-C77]. We analyze the local trends 
 which are of the main focus of the technical analysis. In this article we introduced various 
 statistical quantities examining role of single temporal discretized trend or multitude of grouped 
 trends corresponding to different time delays. Our specific analysis based on 
 Euro-dollar currency pair data at the one minute frequency suggests the importance 
 of cumulative nonrandom effect of trends on the forecasting performance.

 \end{abstract}

 \maketitle

 \section{Introduction}\label{sec:introd}

 The trend extrapolation forecasting is probably most important concept often used by {\em foreign exchange}~(FX) traders.  
 However, the technical analysis is usually rejected because of the {\em efficient market hypothesis} \cite{Fama1970} 
 and its various forms. Assuming validity of the naive trend-following~(TF), one may predict future price 
 that is in line with trend perceived from the historical price records. Clearly, TF does not act 
 in the market alone, it almost always occurs in combination with other 
 (contrarian, fundamentalist) strategies~\cite{Sansone2007}.  

 The concept of trend may represent some practical way for indication, parametrization or quantification 
 of the non-randomness. The analysis of TF efficiency may be applied to measure deviations of the time-series 
 from randomness.  When the trend line is subtracted from the original signal, the residuals may be treated 
 as stationary detrended data. Various non-parametric trend-removal approaches for trend estimation in the presence of the fractal noise 
 have been proposed and applied in~(see e.g.\cite{Afshinpour2008, Chandler2011}).  

 The comparison of the random walk oriented strategy against TF strategy has been discussed~in~\cite{Stevenson1970}. 
 The work has reported observation that speculative price movements on commodity 
 futures do move in a regular, as opposed to random manner. The inefficiency of FX market
 has been investigated by means of the {\em statistical approximate entropy}~\cite{Oh2007}. 
 Similarly, the {\em permutation entropy} has been used to characterize stock market development~\cite{Zunino2009}. 
 The general aspects of TF trading strategy and related risk factors in the hedge fund 
 activities are described in the work~\cite{Fung2001}. 

 In~\cite{JJames2003} the group of trend satisfying currencies has been selected to optimize 
 the portfolio construction.  Note that in the above-mentioned work the trend is determined through the use of the 
 multiple time frame moving averages~(direction indicators) applied simultaneously to meet an optimality 
 criteria for forecast. Concerning relative complexity of aforementioned methods, in this paper we suggest elementary 
 TF strategy which uses only difference between two currency exchange rates separated by given 
 time span called here {\em time scale}. In this form, TF is accomplished 
 as a single parameter prediction method, where the time scale is the only free parameter.  
 We have analyzed statistics of isolated discretized trends as well as trends embedded 
 into tuples containing list of elements sequentially ordered with increasing time scale.

\section{Preprocessing of trading records}\label{sec:Pre}

	We have used six 
        years statistics (2004-2009) of EUR/USD currency exchange rates.  
        The original OANDA broker database of tick-by-tick quotations have been
        transformed into evenly spaced 1 minute time series. 
        The statistics of agents' 
        population proceeds over the {\em ask} currency rate records.

 \subsection{Currency moves made discrete}

 \subsection{The discrete currency exchange rate trends}

\begin{figure}[H]
	\centering
	\includegraphics[width = 0.8\textwidth]{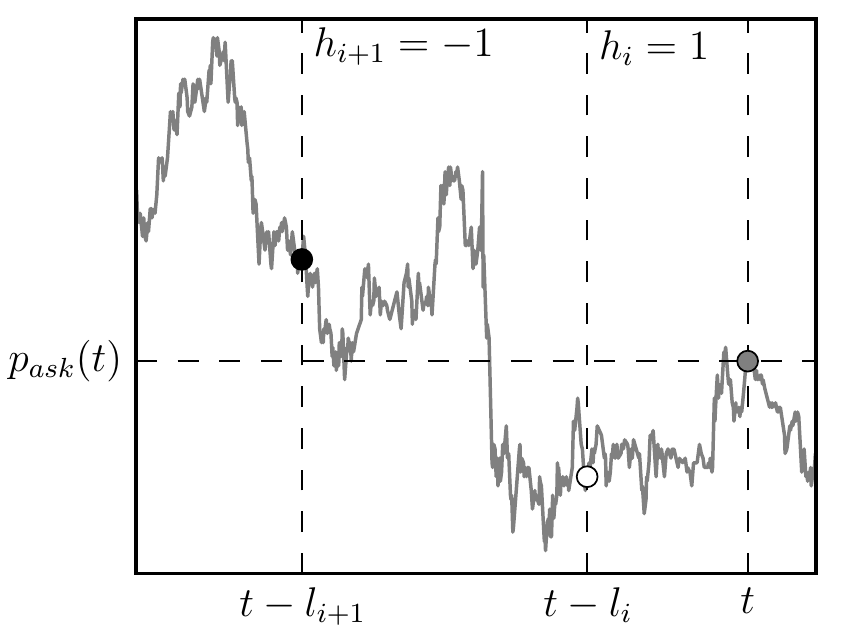}
	\caption{The scheme illustrates specific definition 
       Eq.(\ref{eq:trend1}) of the discrete trend variable 
       $h_i(t)$ regarding subtraction of he actual exchange 
       rate $p_{\rm ask}(t)$ with foregoing $p_{\rm ask}(t - l_i)$. 
       The result is encoded by $-1$  or by $1$.  
       In the sense of remark in subsection~\ref{sec:age}, 
       the situation corresponds to pair of agents operating on the scales $l_i$, $l_{i+1}$.}
	\label{agents's_decission}
        \end{figure}

      The trend is defined by simplest possible manner 
      by comparison of actual currency exchange rate 
      with preceding one. For $i-$th time 
      lag scale $l_i$ and time $t$ we define 
      \begin{eqnarray}
       h_i(t) &=&
       \left\{ 
	\begin{array}{rc} 
	+ 1  \,\,,  &   \,\,  p_{\rm ask}(t)  >     p_{\rm ask}(t - l_i), \\
	- 1  \,\,,  &   \,\,  p_{\rm ask}(t)  \leq  p_{\rm ask}(t - l_i)
	\end{array} 
	\right.\,
	\label{eq:trend1}
      \end{eqnarray}
      where, $p_{\rm ask}(t)$ is the ask rate at which the price-maker 
      is willing to sell 
      the currency.
      Specifically, the time scale may be defined recurrently by
      \begin{eqnarray}
      l_1 = 1\,,\,  
      l_2 = 2\,, \qquad 
      l_i = l_{i-2} + i\,,\,  
      \qquad 
      i = 3, 4,\ldots, N\,.
      \end{eqnarray}
      The advantage of the above choice is that $N=100$ allows 
      nonuniform coverage of the broad domain 
      from the 1 minute to approximately two days with denser 
      occupation of smaller scales.  Clearly, other combinations for $l_i$ are possible.  
      Such freedom of choice evokes question of the relevance and generality 
      of conclusions we reached. Alternative $l_i$ sequences 
      are planned to be studied in a more detail elsewhere.  
      The schematic view on the situation is depicted 
      in Fig.~\ref{agents's_decission}.

      \subsubsection{Multi-trend tuple, pattern with participation of many scales}

      Technical analysis is primarily concerned with identifying patterns of
      prices that repeat themselves~\cite{Zunino2009,Miller1990,Shmilovici2009}. 
      They are mostly based on the symbolic language (lossy encoding) 
      to capture price moves. 
      In further sections we evaluate consequences 
      of the use of N-tuples of primitive discretized trends $h_i(t)$ collected to form 
      \begin{equation} 
      {\bf h}(t)=  \left[\, h_1(t), h_2(t), \ldots, h_N(t)\,\right]\,. 
      \end{equation} 
      This multi-scale~TF structures may be eventually seen as specific patterns 
      of alternating type $-1$ and type $+1$ repeats. 
      Their prediction 
      potential will be studied in the next.

      \subsubsection{Viewing trends as information inputs of autonomous agents}\label{sec:age}

      The tuple ${\bf h}(t)$ may be alternatively viewed as instantaneous 
      macrostate variable for the appropriate decision making of the 
      autonomous agents. In the present approach, each agent 
      is provided with the information 
      about $h_i(t)$
      she/he extracts by measuring the change 
      of the currency rate on its internal specific 
      time scale $l_i$. 

      \subsection{The rate changes accounting for bid/ask spread}

\begin{figure}[H]
	  \centering
	  \includegraphics[width = 0.4\textwidth]{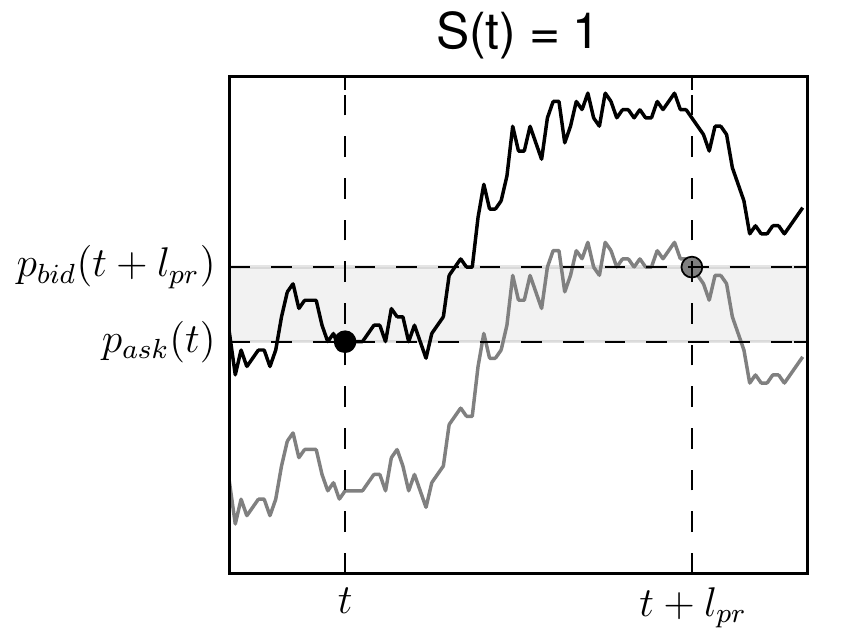}
	  \includegraphics[width = 0.4\textwidth]{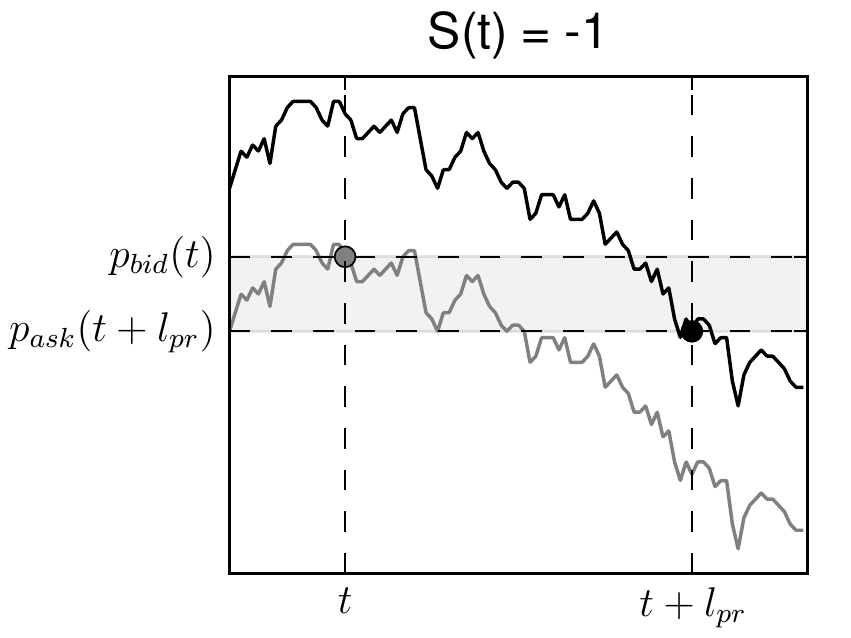} 
	  \includegraphics[width = 0.4\textwidth]{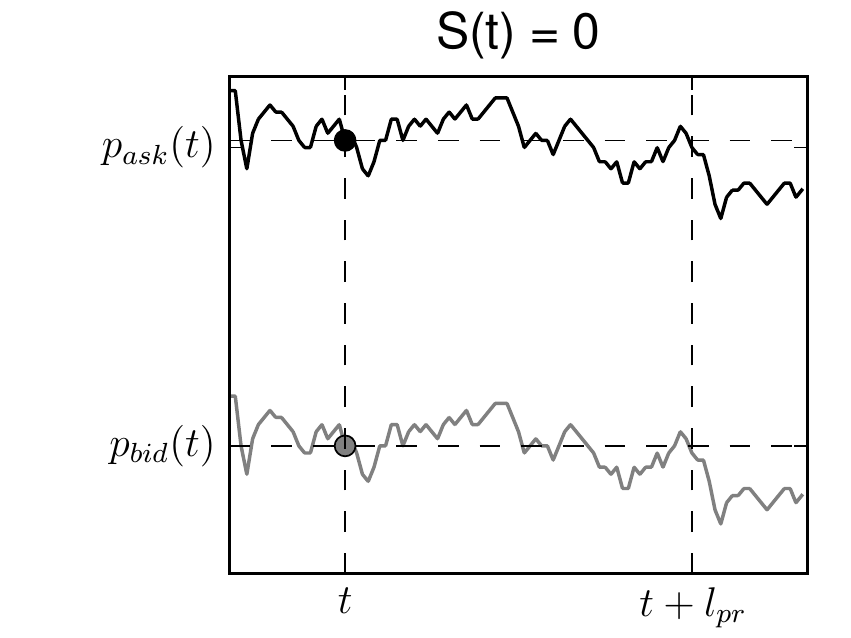}
	  \caption{Figure illustrates discretization - currency rate
          shift $S(t)$ which may 
          attain values $\{-1,0,1\}$ corresponding to 
          different situations as described 
          by Eq.(\ref{eq:spread1}).}
         \label{price_shift_figure}
 	 \end{figure}

	From the FX trader's point-of-view, the currency changes may lead to three principally 
        distinct situations which we suggest to be encoded by 
        three-valued variable $S \in \{-1,0,1\}$ we call {\em currency} {\em rate} {\em shift}. 
        It is defined using discretization of exchange rate as follows   
	\begin{eqnarray}
	 S(t) &=& 
	  \left\{ 
	  \begin{array}{rcl} 
	    -1    &,  &   p_{\rm ask}(t + l_{\rm pr})   <   p_{\rm bid}(t)\,, \\
	    +1    &,  &   p_{\rm bid}(t + l_{\rm pr})   >   p_{\rm ask}(t)\,, \\
	     0    &,  &  \mbox{otherwise}
	 \end{array} 
	 \right.\,
         \label{eq:spread1}
	 \end{eqnarray}
	  where $p_{\rm bid}(t)$ is the demand price of particular 
          currency and $l_{\rm pr}$ is the time horizon for which the currency rate change is being transferred
          to its discrete form. The situation is depicted in Fig.~\ref{price_shift_figure}. The position is supposed to be held from 
          time $t$ to $t + l_{\rm pr}$. For example, in the case of short selling, the holding period $l_{\rm pr}$
          refers to the time between borrowing a currency pair from brokerage, selling it to someone else 
          and then buying it back later. A steep enough currency movement characterized 
          by $|S(t) |=1$ needs to choose appropriate FX operation. It allows the holder of the short position 
          to earn profit from the sale ($S(t)=-1$) or from buying ($S(t)=1$).  
          Clearly, relatively small currency changes are converted to $S(t)=0$.  In this case, 
          the position opening (at $t$) causes the loose because of the bid-ask spread. 
          However, an important issue not discussed in the present work 
          is the number of units in the trade. 

     \section{Statistical analysis of data}

     \subsection{The correspondence between single trend and $S$}

\begin{figure}[H]
	\centering
	{\small (a)}~\includegraphics[width = 0.5\textwidth]{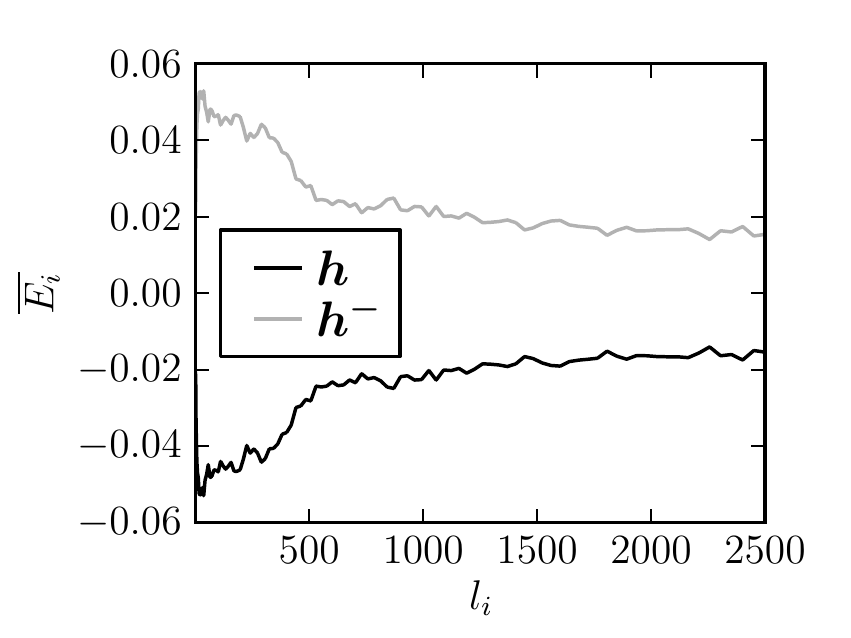}
	{\small (b)}~\includegraphics[width = 0.5\textwidth]{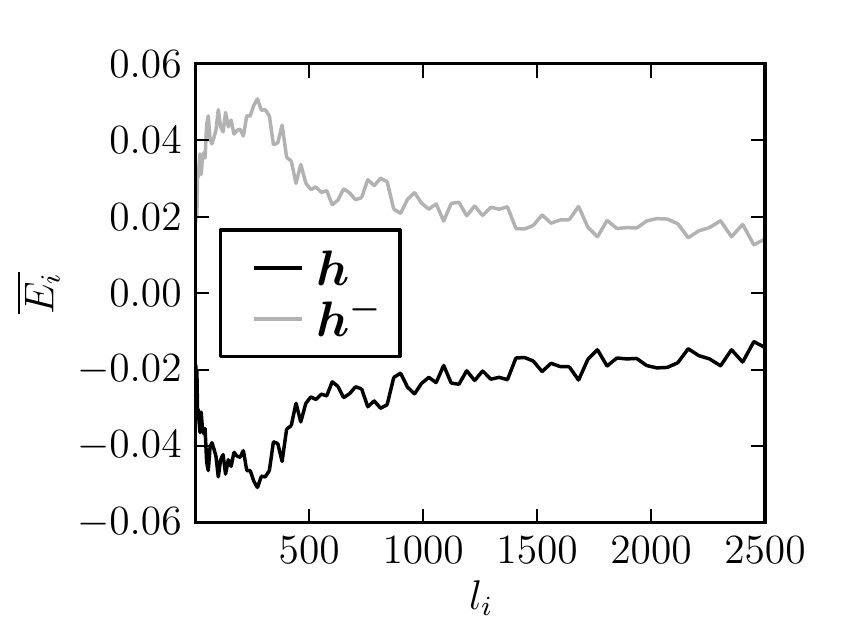} 
	{\small (c)}~\includegraphics[width = 0.5\textwidth]{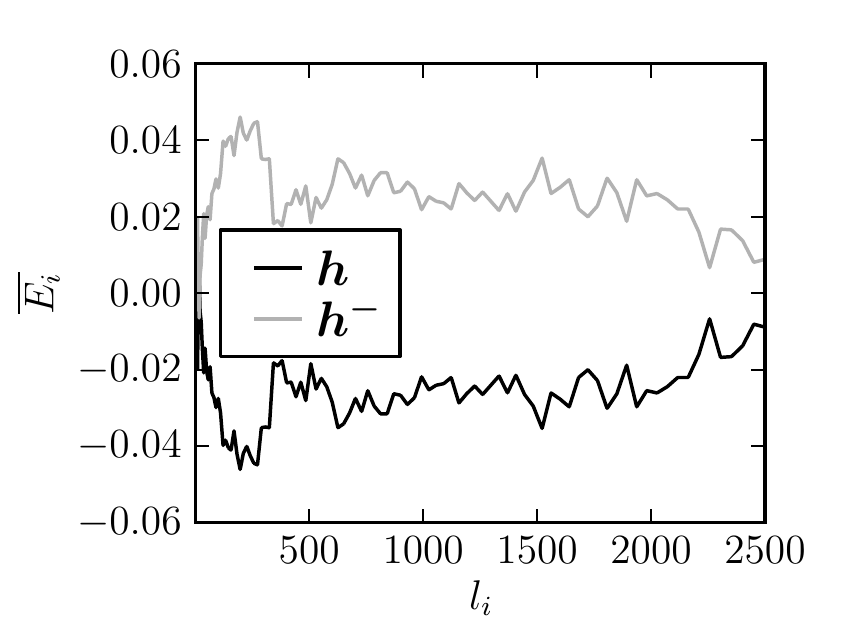}
	\caption{
      Figure showing $\overline{E}_i$ dependence 
      on the time scale $l_i$ obtained 
      for (a)~$l_{\rm pr} = 15$, (b)~$l_{\rm pr} = 60$, and (c)~$l_{\rm pr} = 240$. 
      Surprisingly, the positive matching is achieved 
      by the {\em contrarian strategy}~(counter-trends).}
      \label{mean_matching_figure}
      \end{figure}

      In this section we attempt to answer the question of the 
      relevant time scales 
      for which TF agents come closer to the correct prediction. 
      We came with the proposal of the following conditional 
      statistical {\em matching average} 
      \begin{eqnarray}
      \overline{E}_i =  
      \frac{1}{\#\{t, S(t) \ne 0 \}} \, 
      \sum_{t| S(t) \neq 0}  
      \tilde{\delta}_{[\,h_i(t), S(t)]\,}\,, 
      \end{eqnarray}
      where $\tilde{\delta}_{[.,.]}$ is 
      defined for two arguments as follows  
      \begin{eqnarray}
	\tilde{\delta}_{[A, B]} &=&
	\left\{
	\begin{array}{lcl}
	  +1 & ,  &   A  =   B\,  \\
	  -1 & ,  &   A \ne  B\,  \\
	\end{array}
	\right.\,.
      \end{eqnarray}
     Note that measure identifies the coincidence of two-valued $h_i(t)$ with a non-zero states of $S(t)$. 
     By measuring of $\overline{E}_i$ over the given data 
     for both $h_i(t)$ and counter-trend $h_i^{-}(t)=-h_i(t)$ 
     and their corresponding tuples
     \begin{equation} 
     {\bf h}^{-}(t)=-{\bf h}(t)  
     \end{equation} 
     we obtained $\overline{E}_i$ dependencies on $i$, depicted in Fig.~\ref{mean_matching_figure}. 
     The testing of the effect of the counter-trend expresses 
     that current trends can be broken at any time because 
     of arrival of new market information. Surprisingly, when $\overline{E}_i$ is calculated  
     for counter-trend the matching becomes larger compared to the 
     matching average calculated for ${\bf h}(t)$.

     \subsection{Multi-trend tuple and consequent $S$}

\begin{figure}[H]
	\centering
	\includegraphics[width = 1.0\textwidth]{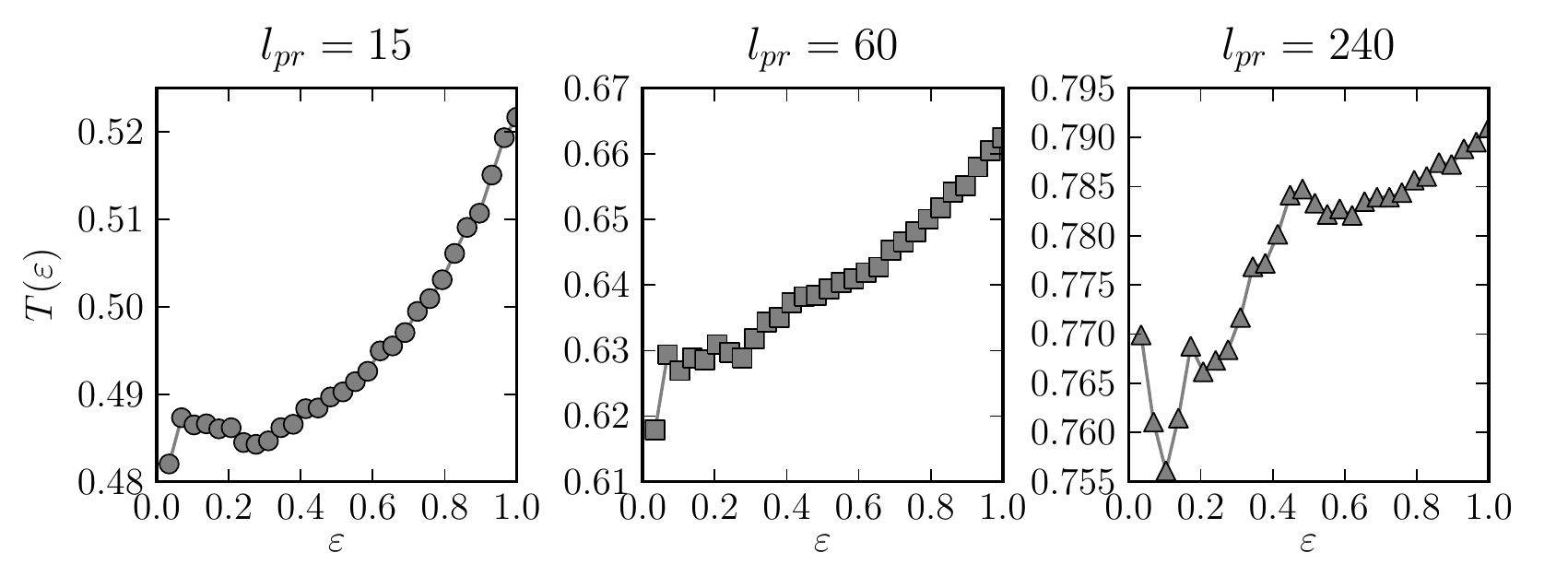}
	\caption{
        The $T(\varepsilon)$ dependence calculated for different 
       time horizons.}
	\label{collective_decision_figure}
        \end{figure}

      In the previous subsection, we studied relationship 
      of single trend with $S(t)$. The logical extension of former efforts 
      will be the study of the correspondence between "action" 
      ${\bf h}(t)$ and "response" $S(t)$. The question arises if the 
      compatibility of many trends (their signs) makes the correspondence 
      more pronounced and satisfying. 
      Here we choose to be less demanding with respect of sign 
      by focussing on the correspondence between the overall 
      effect of trends quantified by the absolute value of instant mean 
      \begin{eqnarray}
      H(t) = \frac{1}{N}\, \Big \vert \sum_{i = 1}^N h_i(t) \Big  \vert\, 
      \end{eqnarray}
      and $|S(t)|$. The degree of $H(t)$, $|S(t)|$ 
      correspondence may be 
      measured by the conditional average
      \begin{eqnarray}
      T(\varepsilon) = \frac{1}{ \# \{t, H(t) < \varepsilon\}  }\,
      \sum_{t | H (t) <  \varepsilon}  \vert S(t) \vert \,,   
      \end{eqnarray}
      where $\#$ denotes the cardinality of the set of events 
      selected according to threshold $H(t)<\varepsilon$. 
      The variable $\varepsilon$ is introduced to control how much homogeneity is in the trends
      $h_i(t)$, $i=1,2, \ldots, N$. Fig.~\ref{collective_decision_figure} shows 
      the cumulative effect represented by $T(\varepsilon)$. 
      As we can see, the monotonicity of $T(\varepsilon)$  
      detectable for $l_{\rm pr} = 15, 60 $ min becomes interrupted at $l_{\rm pr} = 240$ min, where 
      peak is observed.  It seems rather counterintuitive that function $T(\varepsilon)$ 
      is increasing in $l_{\rm pr}$. This fact stems from the relative suppression 
      of $S=0$ states at larger $l_{\rm pr}$, where currency changes become excessive.   

       \subsection{Similarity of multi-trend tuples and predictability}

\begin{figure}[H]
       \centering
       {\small (a)}~\includegraphics[width = 0.91\textwidth]{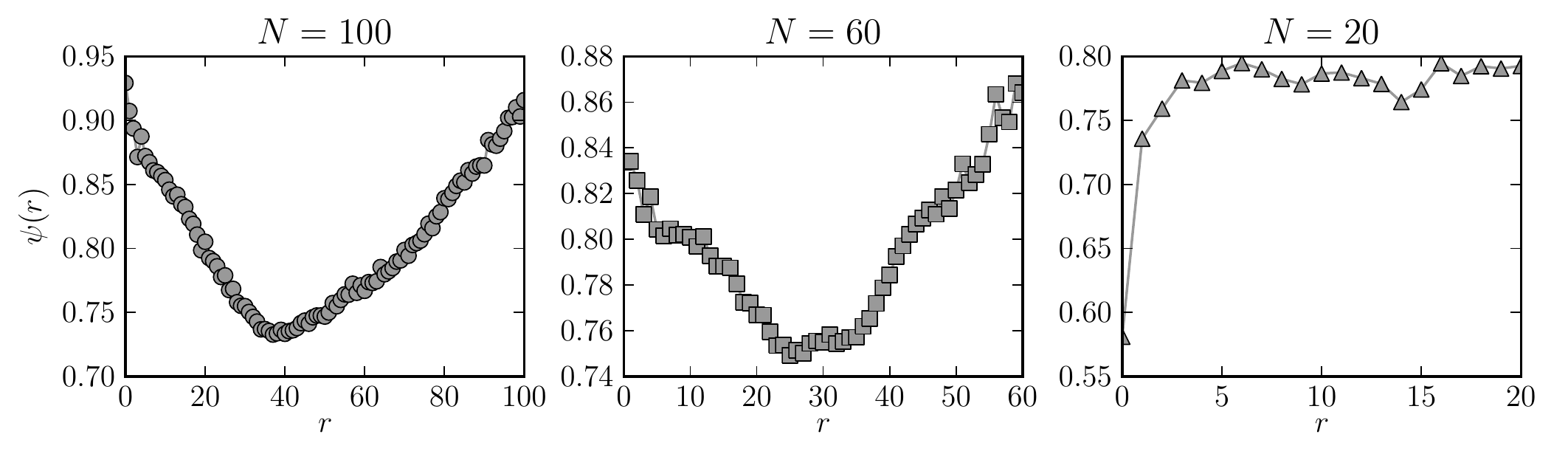}
       {\small (b)}~\includegraphics[width = 0.91\textwidth]{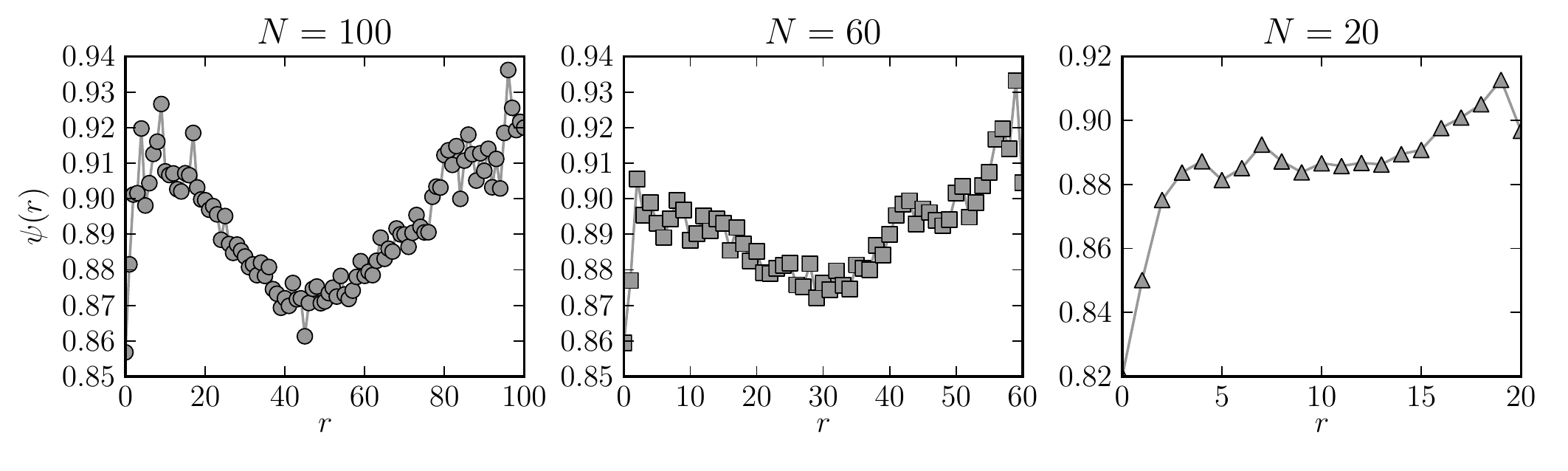}
       {\small (c)}~\includegraphics[width = 0.91\textwidth]{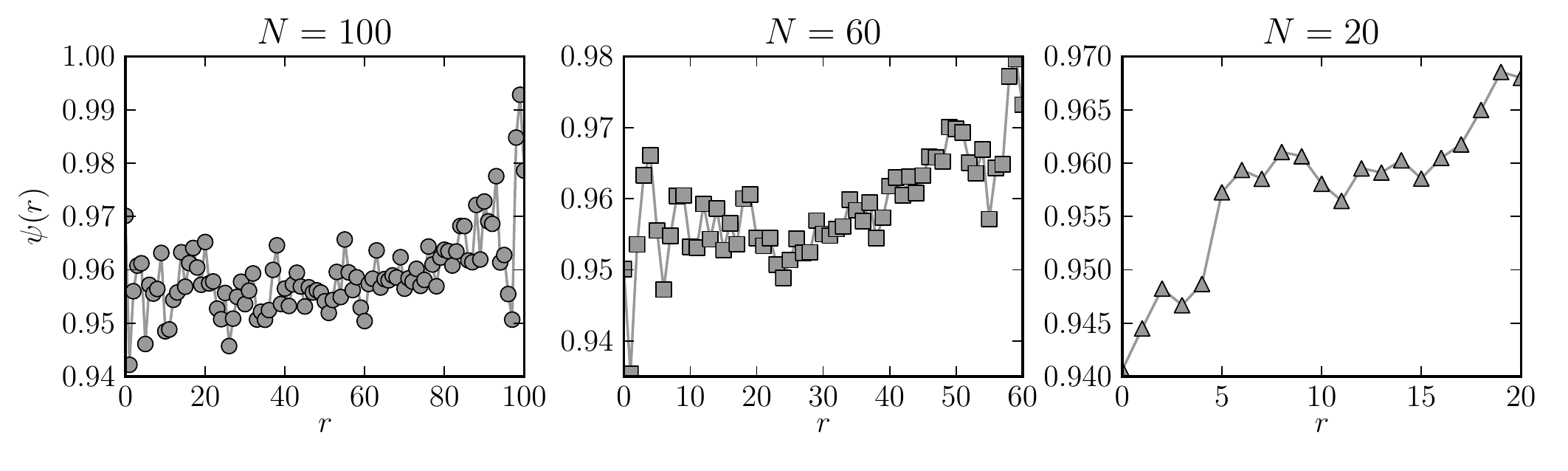}
	\caption{$\psi(r)$ dependence calculated for different time horizons:~(a)~$l_{\rm pr}=15$ min; 
        (b)~$l_{\rm pr}=60$ min; (c)~$l_{\rm pr}=240$ min. 
        Among the selected cases, the $N=20$ tuple with monotonously 
        increasing $\psi(r)$ is roughly the most perspective 
        for the forecasting purposes.} 
	\label{pattern_comparison_figure}
        \end{figure}

      In this subsection we introduce measure which admits to analyze prediction 
      abilities based on the trend tuples. 
      The basic idea behind can be seen as continuity
      in the multivariate space of dimension $N$. 
      The dissimilarity between two trend tuples ${\bf h}(t_i)$ and ${\bf h}(t_j)$ measured 
      at different moments $ t_i, t_j$ may be expressed by means of 
      {\em Hamming-like distance}
      \begin{eqnarray}
      D_{{\rm h},(i, j)}  =   
      \sum_{k=1}^N \left(\, 1 -  \delta \left(\, h_k(t_i), h_k(t_j) \right)\,\right)\,, 
      \end{eqnarray}
      where $\delta(x,x)=1$, and $\delta(x,y)=0$ as $x\neq y$. 
      Then the statistics may be represented by the histogram 
      of $|S(t_i)-S(t_j)|$ differences 
       \begin{eqnarray}
        \psi(r) = 
	\frac{1}{ \# \{i, j, D_{{\rm h},(i, j)} = r\,\} }\,
	\sum_{i,j |\, D_{{\rm h},(i, j)} = r} \vert S(t_i)  -  S(t_j) \vert\,.  
      \end{eqnarray}
      Note that summation is carried out only for pairs that satisfy constraint
      $D_{{\rm h},(i,j)}=r$. 
      The assumption of continuity implies monotonic increase of $\psi(r)$ on the 
      $r$ (related to Hamming-like distance) as $r$ 
      is sufficiently small. Other words, similarity of the tuples 
      should imply similarity in the currency 
      rate shifts~(see Fig~\ref{pattern_comparison_figure}).  
      The data analysis show us that monotonicity takes place in the
      limited $N$, $l_{\rm pr}$ parametric region only, while the 
      combination of large $l_{\rm pr}$ 
      with small $N$ better achieves the expected monotonicity 
      of $\psi(r)$. 
      
      We may conclude from this that character of FX data and trading data
      in general impose restrictions on the overall tuple size and only 
      tuples of the restricted size may be efficient in the reducing 
      of the forecast error $\sim \vert S(t_i) - S(t_j) \vert$. Results 
      obtained show that the best performance is achieved within the range 
      $12<N<20$~(only three $N$ are plotted due to lack of space). 

    \section{Conclusions}
   
    We investigated statistical properties of TF and multi-TF strategies. 
    We started with elementary TF forms where information comes 
    solely from the difference of two prices (two currencies) separated by the fixed time
    scale~(delay).  The study of statistics revealed clear non-suitability of naive trend 
    estimator for the forecast purposes. According to the results obtained, 
    the trend continuation is less good compared to the contrarian 
    strategy of the trend turnover. Similar and quite universal conclusion 
    has been confirmed by the previous studies performed for different market data 
    (including e.g. commodities) which span over different time scales.

    The statistical analysis supported development and implementation 
    of parallelized TF strategies build upon the tuple which consists  
    of many trends for contributing time scales.  We have shown 
    that one may benefit from the prediction potential of such complex information 
    structures, however, the vast majority of problems remain to be solved. The 
    optimal inter-tuple distance in common with suitable multi-valued 
    trend discretization which regards bid-ask spread could 
    be essential ingredients required for success of such efforts. 
    In such formulation nearly optimal $N$ may play the role of the 
    {\em embedding} {\em dimension} \cite{Grassberger1983}, 
    i.e. minimum number of independent variables necessary 
    to describe system. Intuitively, the parameter 
    $N$ optimized to produce most reliable forecast 
    (including optimization of $l_{\rm pr}$ scale) 
    may be different from the value of 'true' embedding dimension.  

    In the future, it would be interesting to experiment 
    with {\em recurrence quantification} {\em analysis}~\cite{Marwan2008,Bigdeli2009} 
    to characterize complexity in iterations of the multi-trend 
    tuples (patterns) obtained 
    from the time series 
    of generally non-stationary market rate returns.

\end{document}